\documentclass[aps,reprint,amsmath,amssymb,showkeys,superscriptaddress]{revtex4-2}

\usepackage{graphicx}
\usepackage{siunitx}
\usepackage{comment}
\usepackage{physics}
\newcommand{\scomment}[1]{}
\begin{document}

\title{Self-consistent solution of magnetic and friction energy losses\\ of a magnetic nanoparticle}

\author{Santiago Helbig}
\email{santiago.helbig@univie.ac.at}
\affiliation{University of Vienna, Faculty of Physics, Physics of Functional Materials, Kolingasse 14-16, 1090 Vienna, Austria}
\affiliation{University of Vienna, Vienna Doctoral School in Physics, Boltzmanngasse 5, 1090 Vienna, Austria}
\author{Claas Abert}
\affiliation{University of Vienna, Faculty of Physics, Physics of Functional Materials, Kolingasse 14-16, 1090 Vienna, Austria}
\affiliation{University of Vienna, Faculty of Physics, MMM Mathematics-Magnetism-Materials, Oskar-Morgenstern-Platz 1, 1090 Vienna, Austria}
\author{Pedro A. S\'{a}nchez}
\affiliation{University of Vienna, Faculty of Physics, Computational and Soft Matter Physics, Boltzmanngasse 5, 1090 Vienna, Austria}
\affiliation{University of the Balearic Islands, Physics Department, 07122 Palma de Mallorca, Spain}
\author{Sofia S. Kantorovich}
\affiliation{University of Vienna, Faculty of Physics, Computational and Soft Matter Physics, Boltzmanngasse 5, 1090 Vienna, Austria}
\affiliation{University of Vienna, Faculty of Physics, MMM Mathematics-Magnetism-Materials, Oskar-Morgenstern-Platz 1, 1090 Vienna, Austria}
\affiliation{Ural Federal University, Institute of Mathematics and Natural Sciences, Lenin av. 51, 620000 Ekaterinburg, Russian Federation}
\author{Dieter Suess}
\affiliation{University of Vienna, Faculty of Physics, Physics of Functional Materials, Kolingasse 14-16, 1090 Vienna, Austria}
\affiliation{University of Vienna, Faculty of Physics, MMM Mathematics-Magnetism-Materials, Oskar-Morgenstern-Platz 1, 1090 Vienna, Austria}

\date{\today}

\begin{abstract}
	We present a simple simulation model for analysing magnetic and frictional losses of magnetic nanoparticles in viscous fluids subject to alternating magnetic fields. Assuming a particle size below the single-domain limit, we use a macrospin approach and solve the Landau-Lifshitz-Gilbert equation coupled to the mechanical torque equation. Despite its simplicity the presented model exhibits surprisingly rich physics and enables a detailed analysis of the different loss processes depending on field parameters and initial arrangement of the particle and the field. Depending on those parameters regions of different steady states emerge: a region with dominating N\'eel relaxation and high magnetic losses and another region region with high frictional losses at low fields or low frequencies. The energy increases continuously even across regime boundaries up to frequencies above the Brownian relaxation limit. At those higher frequencies the steady state can also depend on the initial orientation of the particle in the external field. The general behavior and special cases and their specific absorption rates are compared and discussed.
\end{abstract}

\keywords{Magnetic nanoparticle, Magnetic Hyperthermia, Numerical Simulations}
\maketitle

\section{Introduction}\label{intro}
The versatile properties of magnetic fluids are very attractive for applications in biomedicine, for example as contrast agents, in drug targeting or for hyperthermia \cite{2015ApPRv...2d1302P,pollert2013magnetic,2003JPhD...36R.167P,KUMAR2011789,doi:10.1080/02656730802104757,ObaidatIhabM2015MPoM}. The magnetic characteristics of the fluid stem from the suspended single-domain magnetic nanoparticles (MNP).
Depending on the requirements for an application the material properties have to be chosen carefully and the field parameters need to be tuned for optimal control and efficiency. For magnetically induced hyperthermia in cancer treatment the requirement is to generate as much heat as necessary to destroy cancer tissue.
For this procedure, a magnetic fluid suspension is injected intra-tumoral or close to the cancer cells. An external field gradient can be used to guide and focus the magnetic fluid closer to the cancer cells. Then an alternating magnetic field (AMF) can be used to induce heat in the magnetic fluid by stimulating the MNPs and locally destroy the cancer cells \cite{https://doi.org/10.1002/advs.202105333, doi:10.1080/02656730802104757}.
The heat induced inside an MNP by a magnetic AC-field is explained by two processes, the Brownian relaxation \cite{BrownWilliamFuller1963M} and N\'eel relaxation \cite{neel1949theorie}, and is generated by surface friction and internal switching of the magnetization.
Both processes occur simultaneously and by analyzing the energy losses due to the viscous torque and the hysteresis curves, the heating of the single-domain MNP in a viscous fluid can each be individually quantified with the simulation model presented in this work.

Studying the heating properties of MNPs with experiments \cite{doi:10.3109/02656736.2013.822993,AtkinsonWilliamJ1984UFiH,dutz2007hysteresis} and simulations \cite{2015NatSR...5E9090R,HovorkaO2014Rogs,doi:10.1063/1.3551582,L.RaikherYu2014Paom} is a very active field of research.
Usadel \textit{et al.} \cite{doi:10.1063/1.4937919} developed a numerical approach with a system of kinetic equations. They found that the behavior of a particle depends on the field parameters and found two steady states in the zero-temperature case. Furthermore, the influence of temperature on the system is a major point in that work.

In contrast, in our work the energy dissipation of a magnetic particle under athermal conditions is studied. Although the thermal fluctuations of the system have been omitted, the self-consistent solution of the coupled magnetization and mechanical dynamics exhibit rich physics. A large parameter study of the field is conducted and dependencies are discussed in a comprehensive fashion. The influences and individual contributions of N\'eel and Brownian relaxation are analyzed in detail. The emerging two steady states depend on the dominating relaxation process and lead to turning of the particle or allow for the switching of the magnetization.
Moreover, while the two steady states can be separated into two regions in the parameter space of field strength and frequency, we also observed a third region depending on the initial orientation of the particle's easy axis relative to the field axis.

In section \ref{method} the model and methods to calculate the energy losses as well as other comparison models are introduced. The results of the different models and interpretation of such will be discussed in section \ref{results}. A conclusion is drawn in the last section \ref{conclusion}.

\section{Model}\label{method}
Our system consists of a single-domain MNP in a viscous carrier fluid. The particle is assumed to be spherical with an uniaxial crystalline anisotropy. This effectively gives the particle, although spherically symmetrical, a preferred axis of the magnetic moment which is also referred to as the anisotropy axis or easy axis. In the absence of any external influence, the magnetization is relaxed and resides in the easy axis of the particle. When applying an external magnetic field that is not aligned with the easy axis, the magnetic moment of the particle will decouple from the easy axis in a manner that is defined by the Landau-Lifshitz-Gilbert (LLG) equation \cite{landaulifshitz1935,gilbert1955lagrangian,gilbert2004phenomenological}.
The misalignment of the magnetization and the easy axis exerts a mechanical torque rotating the particle. This effect is called Brownian relaxation and will lead, again, to an alignment of the magnetic moment and easy axis with the external field after some settlement time.

Unless another force is acting on the particle, its rotation is limited to a 2D plane, spanned by the easy axis and the field axis, with the direction of the magnetic moment somewhere in between. In three dimensions the 2D rotation for random orientations of the easy axis will be symmetric around the field axis. Hence, the whole system can be described by the angles $\theta$, for the direction of the magnetization, and $\phi$, the orientation of the easy axis of the particle, see Fig. \ref{fig:astr}. In order to model the rotation in a plane, a system of equations of torques is constructed. Three torques are considered: the magnetic, viscous and inertial torque. Due to the collinearity of the torques, they can be reduced to their scalar value in the following equations.

The driving torque of the system is the magnetic torque $\tau_\text{mag}$ exerted by the magnetic field
\begin{equation}\label{eq:tmag}
	\tau_\text{mag} = \mu_0 M_\text{s} V_\text{m} H \sin (\theta)
\end{equation}
with the vacuum permeability $\mu_0$, the material specific saturation magnetization $M_\text{s}$, the magnetic volume of the particle $V_\text{m}$, the applied field intensity $H$ and the angle $\theta$, describing the angle between the magnetization and the field.
In this work, the field $H$ is an alternating field.

The viscous torque $\tau_\text{visc}$ opposes the magnetic torque and decelerates the rotation. The standard hydrodynamic result for viscous torque of a spherical object is given by
\begin{equation}\label{eq:tvisc1}
	\tau_\text{visc} = - 8 \pi r^3 \eta \; \dot{\phi} = - 6 V \eta \; \dot{\phi} \text{,}
\end{equation}
with the hydrodynamic radius of the particle $r$ and the hydrodynamic volume $V$, the dynamic viscosity of the carrier fluid $\eta$ and the angular velocity $\dot{\phi} = \dv{\phi}{t}$ of the easy axis of the particle and denotes the change in the angle $\phi$ between the easy axis and the field. The minus indicates the opposition to any driving torque.

The third torque acting on the particle is the inertial torque $\tau_\text{inert}$ of a spherical object
\begin{equation}\label{eq:tinert1}
	\tau_\text{inert} = \frac{2}{5} m_\star \: r^2 \ddot{\phi} = \frac{2}{5} \rho \; V r^2 \ddot{\phi} \text{.}
\end{equation}
In this equation $m_\star$ refers to the mass of the particle. It is denoted with a star and rewritten in order to highlight the dependence on the volume and to avoid any confusion with the magnetic moment $\boldsymbol{m}$.
$\ddot{\phi} = \dv[2]{\phi}{t}$ denotes the angular acceleration of the easy axis.

The mechanical equation of motion can be derived from the conservation of angular momentum: $\tau_\text{inertia} = \tau_\text{mag} + \tau_\text{visc}$. Inserting from Eq. \eqref{eq:tinert1},\eqref{eq:tvisc1} and \eqref{eq:tmag} yields the angular acceleration
\begin{equation}\label{eq:com}
	\ddot{\phi} = \frac{5}{2} \frac{\tau_\text{mag} + \tau_\text{visc}}{\rho V r^2} .
\end{equation}
For the magnetic equation of motion, we consider the Landau-Lifshitz-Gilbert equation. Since the frequency of the external field is considered to be small compared to the characteristic magnetic precession frequency, the LLG is reduced to the damping term yielding 
\begin{equation}\label{eq:rllg}
	\dot{\boldsymbol{m}} = - \frac{\alpha \gamma}{1 + \alpha^2} \; \boldsymbol{m} \times (\boldsymbol{m} \times \boldsymbol{H}_\text{eff}),
\end{equation}
where $\alpha$ is the Gilbert damping constant, $\gamma$ denotes the reduced gyromagnetic ratio and $\boldsymbol{H}_\text{eff}$ is the effective field and in this case consists of the anisotropy field $\boldsymbol{H}_\text{ani}$ and the external field $\boldsymbol{H}$. The anisotropy field is chosen for uniaxial crystalline anisotropy \cite{kronmuller2007general}:
\begin{equation}\label{eq:ani}
	\boldsymbol{H}_\text{ani} = \frac{2 K_u}{\mu_0 M_s} \left( \boldsymbol{m} \cdot \boldsymbol{n} \right) \boldsymbol{n}
\end{equation}
Although the particle rotates in a plane, for the numerical time integration the polar coordinates are transformed into three dimensional Cartesian coordinates for broader application with the angular velocity $\dot{\boldsymbol{\phi}} = \boldsymbol{n} \times \dot{\boldsymbol{n}}$ and angular acceleration $\ddot{\boldsymbol{\phi}} = \boldsymbol{n} \times \ddot{\boldsymbol{n}}$, where $\boldsymbol{n}$ denotes the orientation of the easy axis.
The coordinate system is chosen such that the field is aligned with the x-axis of the coordinate system, while the easy axis is located somewhere in the x-z-plane. Therefore the angles $\theta$ and $\phi$ correspond not only to the angle relative to the field but also to the x-axis.

A coupled system of equations of the orientation of magnetization (combining Eq. \eqref{eq:rllg} and Eq. \eqref{eq:ani}) and easy axis (combining Eq. \eqref{eq:com} and Eq. \eqref{eq:tmag}, replacing the $\sin(\theta)$ in Eq. \eqref{eq:tmag} with the cross-product $\boldsymbol{m} \times \boldsymbol{H}$) is evolved, resulting in the following two equations:
\begin{equation}\label{eq:rllg2}
	\dot{\boldsymbol{m}} = - \frac{\alpha \gamma}{1 + \alpha^2} \; \boldsymbol{m} \times \left( \boldsymbol{m} \times \left( \frac{2 K_u}{\mu_0 M_s} \left( \boldsymbol{m} \cdot \boldsymbol{n} \right) \boldsymbol{n} + \boldsymbol{H} \right)\scomment{closing bracket was wrongly set, 02.06.22} \right)
\end{equation}

\begin{equation}\label{eq:ddn}
	\ddot{\boldsymbol{n}} = \frac{5}{2} \frac{\mu_0 M_\text{s} V_\text{m} \boldsymbol{m} \times \boldsymbol{H} - 6 V \eta \; \boldsymbol{n} \times \dot{\boldsymbol{n}}}{\rho V r^2} \times \boldsymbol{n}
\end{equation}
For the numerical time integration of the system the following state vector $\boldsymbol{x}$ and its derivative are used:

\begin{align*}
	\boldsymbol{x} = \left[ \begin{matrix}
		\boldsymbol{m} \\
		\boldsymbol{n} \\
		\dot{\boldsymbol{n}}
	\end{matrix} \right] && \dv{t} \; \boldsymbol{x} = \left[ \begin{matrix}
		\text{RHS of eq.} \; \ref{eq:rllg2} \\
		\dot{\boldsymbol{n}} \\
		\text{RHS of eq.} \; \ref{eq:ddn}
	\end{matrix} \right]
\end{align*}
Due to the high stiffness of this system the implicit Runge-Kutta scheme, the Radau solver from the scipy-library, with adaptive time steps was used. The adaptive time steps are important to join together the two different time scales of the mechanical and magnetic dynamics especially the switching of the magnetization.

\subsection{Energy losses}\label{sec:ene}
From the state of the particle, the angular velocity can be determined which allows for the calculation of the viscous torque and the dissipated friction energy per cycle
\begin{equation}\label{eq:e_fric}
	E_\text{fric} = \oint_c \tau_\text{visc} \, \dot{\phi} \; \text{d}t.
\end{equation}
A constant high angular velocity thus maximizes the friction.
Due to the symmetry of the anisotropy and the nature of the AMF, the arc of the rotation is not a full circle but a half circle with acceleration and deceleration as the field alternates. Thus, the particle cannot maintain a constant angular velocity throughout a cycle of the field.
For an AMF the optimal conditions occur when the particle is close to the rotational limit and the particle can remain in motion with short acceleration and deceleration phases.

For magnetic systems, the dissipated energy of the system can be calculated with help of the hysteresis loop. The area of the hysteresis loop over a cycle $c$ times the magnetic volume $V_m$ of the particle results in the dissipated energy
\begin{equation}\label{eq:mag_loss}
	E_\text{hyst} = \mu_0 V_m \oint_c \boldsymbol{M} (\boldsymbol{H}) \; \text{d}\boldsymbol{H} .
\end{equation}
Since the direction of the AMF is fixed in the global frame of refrerence, the total energy losses can be calculated as the scalar integral along over the projection of the magnetization onto the field direction. In our simulations the field axis is aligned with the x-axis.
Transforming the system into the rotating frame of reference of the particle yields the magnetic losses. The calculation is analogous to a rotation of the field and an immobilized particle as described in \cite{della2010identifying,2002JMMM..252..370R}.
The total losses are comprised of the losses due to Brownian and N\'eel relaxation while in the rotating reference frame only the losses due to N\'eel relaxation are captured.

In order to check consistency of our model, we compare the total energy loss computed according to Eq. \eqref{eq:mag_loss} and the sum of magnetic and frictional losses:
\begin{align}\label{eq:tot_loss}
		E_\text{total} \quad& = &E_\text{mag} \quad \;& +&E_\text{fric} \quad \\
		\mu_0 V_m \oint_c M_x \; \text{d}H_x& = &\mu_0 V_m \oint_c \boldsymbol{M}_{\oslash} \; \text{d}\boldsymbol{H}_{\oslash} \;& +&\oint_c \tau_\text{visc} \, \dot{\phi} \; \text{d}t
\end{align}
The subscript $\oslash$ denotes the coordinates in the rotating reference frame of the particle. The numerical difference of the two calculation methods will be discussed in detail later in this work.

We will also use the power dissipation per mass $m_\star$, the specific absorption rate SAR, which is an important measure of heating efficiency for magnetic hyperthermia
\begin{equation}\label{eq:SAR}
	SAR = E_\text{total} \, \frac{f}{m_\star} .
\end{equation}
Here, $f$ denotes the frequency of the AMF and $m_\star$ denotes the mass of the particle.

\subsection{Comparison models}
The hybrid method, which has been developed in this paper, merges two other methods, the immobilized and rigid method.
The hybrid solution is simulated with a non-magnetic surfactant layer, which is necessary for bio-compatibility in medical applications and to prevent aggregation of the particles, additionally the bulk solution, where the whole particle volume is magnetized, is also shown. This leads to a stronger magnetic and reduced viscous torque and thus higher rotation amplitudes.
For the hybrid solutions the magnetization is not strictly bound to the easy axis, but at these small field strengths the decoupling of the magnetization from the easy axis is seemingly minuscule. Still, this small deviation is expected to represent a more realistic scenario.

The immobilized method refers to the immobilization of the easy axis, for example by increasing the viscosity or enclosing the particle in a solid material, which then only allows for motion of the magnetization and thus is a solely magnetic system.
This can be described exactly by the Stoner-Wohlfarth model. The particle's easy axis is fixated in its position and only the magnetization moves relative to the easy axis.
The solution of the Stoner-Wohlfarth model has to match the results of the simulated N\'eel relaxation.

On the other hand, the rigid solution is a mechanical system, where the magnetic moment is strictly bound to the easy axis. Thus the whole particle rotates like a compass needle to align with the magnetic field. The rigid particle problem can also be solved analytically.

The full equation of motion of the hybrid model can only be solved by means of numerical simulations. However, in the limiting case of very high magnetic anisotropy, the system reduces to the mechanical equation of motion which allows for an analytical solution.
Furthermore, we assume the magnetic and viscous torques scale equally with the size of the particle while the inertial torque at this scale is many magnitudes smaller and can be neglected \cite{purcell1977life}. This means that inertial effects stop immediately once a force or torque stops acting on the body.
When the complete particle can be magnetized then $V_m = V$, omitting the surfactant layer, the system of equations can be further simplified to a size- and mass-independent model with only magnetic and viscous torques.
\begin{equation}\label{eq:ana}
	\dot{\phi} \; (t) = \frac{\mu_0 M_\text{s} H \sin({\phi}) \sin({2 \pi f t})}{6 \; \eta} .
\end{equation}
Here, $\dot{\phi}$ denotes the angular velocity and $\phi$ the angle of the easy axis and the magnetization relative to the field.
$H$ is the maximum field amplitude of the external applied field and $\eta$ is the viscosity of the fluid. 
This first order ODE can be analytically solved which results in
\begin{equation}\label{eq:ana2}
	\phi (t) = 2 \cot^{-1} \left(e^{k + \mu_0 M_\text{s} H \cos({2 \pi f t})/{(12 \eta \pi f)}} \right).
\end{equation}
Because the magnetic moment of the particle cannot decouple from its easy axis, one angle $\phi$ is enough to describe the orientation of both easy axis and magnetic moment. The integration constant k can be determined by the initial condition for $\phi$.
\begin{equation}\label{eq:anaconst}
	k (\phi_0) = \ln \left(\tan\left(\frac{\pi-\phi_0}{2}\right)\right) - \frac{\mu_0 H M_s}{6 \eta} \frac{1}{2 \pi f}
\end{equation}
In order to compare different calculation methods the initial angle is chosen to be $\phi_0 = \phi (t=0) = \pi / 2$. This leads to the integration constant
\begin{equation}
    k = \mu_0 M_\text{s} H /{(12 \eta \pi f)} .
\end{equation}

The five different models, the hybrid model with and without a surfactant layer, the rigid model and its analytic solution and the immobilized model will be discussed in the next section.

\begin{figure}[b]
	\includegraphics[width=0.8\columnwidth]{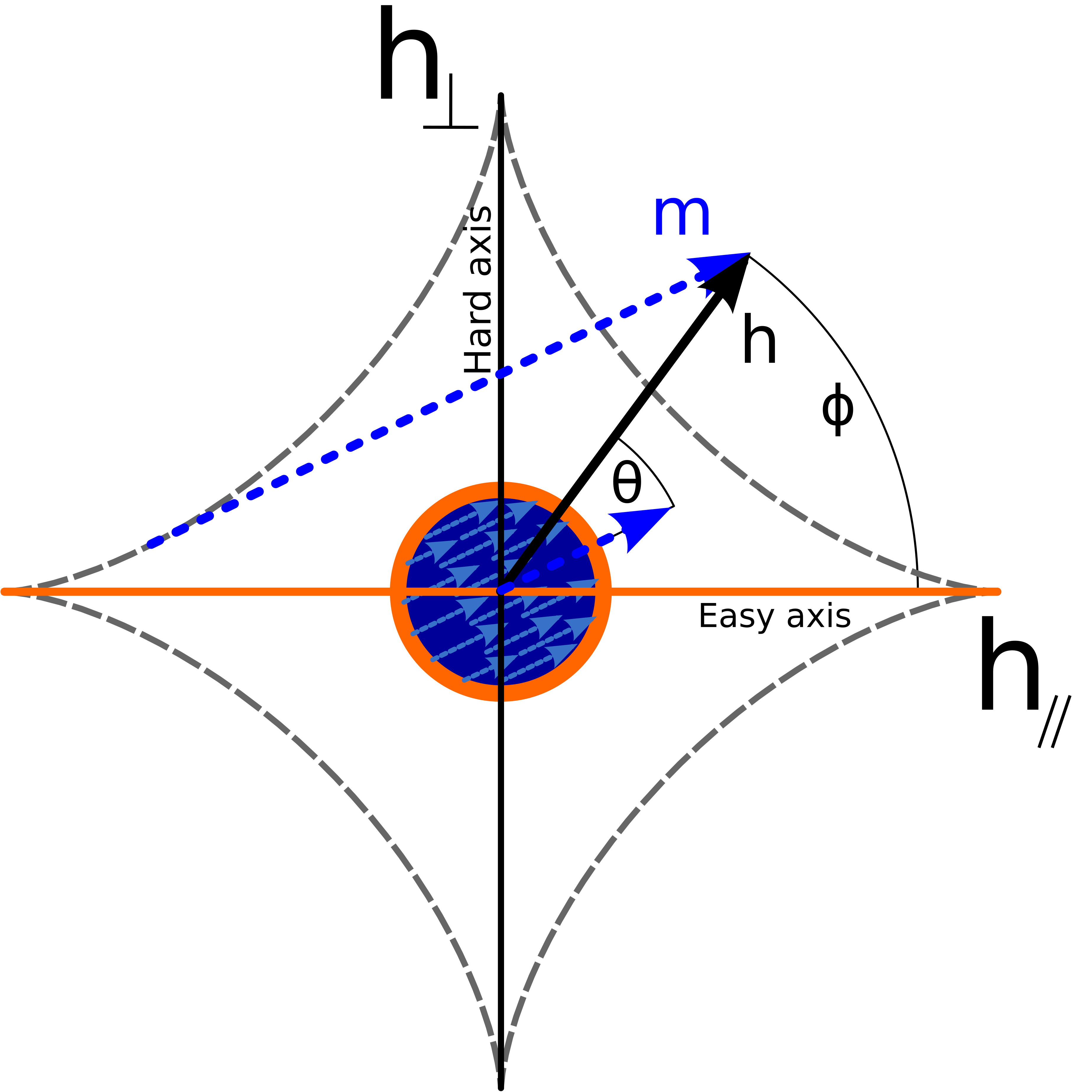}
	\caption{
	The Stoner-Wohlfarth astroid. The orange horizontal line indicates the orientation of the easy axis. 
	The black arrow $H$ indicates the applied field, with an angle $\phi$ relative to the easy axis, while the long blue arrow $M$ draws a tangent on the astroid to the tip of the field vector $H$ which gives the orientation of the magnetic moment of the particle. It is indicated by the short blue arrow, with angle $\theta$ relative to the field, drawn parallel to the long blue arrow at the center.
	}
	\label{fig:astr}
\end{figure}

\section{Results}\label{results}
For simplicity we consider the limit of infinite dilution of the magnetic fluid and simulate one spherical uniaxial single-domain MNP suspended in a viscous fluid. The particle is made out of a magnetite-like metal with a \textit{saturation magnetization} $M_\text{s} = \SI{400000}{A/m}$, a \textit{density} $\rho = \SI{5170}{kg/m^3}$ and an \textit{anisotropy constant} $K_\text{u} = \SI{30000}{J/m^3}$.
This yields a critical switching field of $H_\text{crit} \approx \SI{120}{kA/m}$.
The particle is spherical with a \textit{radius} $r = \SI{9}{nm}$ and an additional \textit{surfactant layer} $hs = \SI{1}{nm}$.
The surfactant layer is considered non-magnetic and mass-less (much less dense than the magnetic core of the particle) and therefore only contributes to the viscous torque by increasing the surface friction and damping the movement of the particle.
The surrounding carrier fluid has the \textit{dynamic viscosity parameter} $\eta = \SI{0.89}{mPa \, s}$ which corresponds to the viscosity of water at \SI{20}{\celsius}. The dimensionless Gilbert \textit{damping parameter} is chosen to be $\alpha = 0.08$. For the simulations the parameters are kept mostly consistent and any deviations from our standard values will be highlighted.
The simulations are run for over 50 cycles of the AMF which should be enough time for the particle to settle in a steady state. The particle settles in fewer cycles in its steady state for lower frequencies.

\begin{figure}
	\includegraphics[width=\columnwidth]{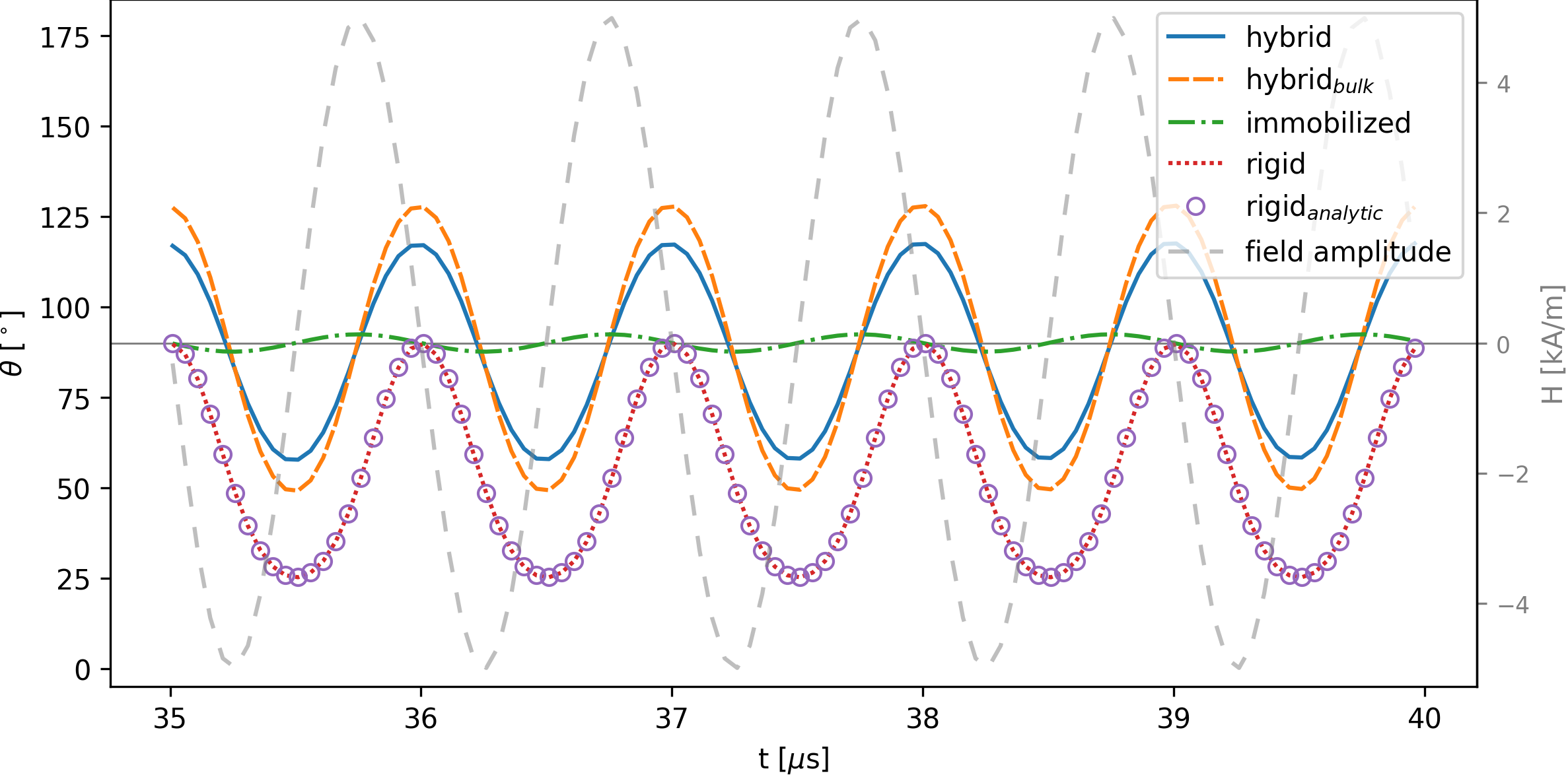}
	\caption{\label{fig:ana}
	Plot of the oscillation of $\theta$ in time.
	Taken for different initial configurations for a particle with \SI{10}{nm} radius, a field strength of \SI{5}{kA/m} and a frequency of \SI{100}{kHz}\scomment{changed from MHz to kHz, typo, 01.06.2022}, the trajectories are compared. Here shown: with decoupled magnetization and a surfactant layer (\textit{hybrid}), when completely magnetized and no surfactant layer (\textit{hybrid}$_\text{bulk}$), for the magnetic system (\textit{immobilized} easy axis), for the mechanical system (\textit{rigid}) and the \textit{analytic} solution. The dashed light gray line indicates the amplitude of the field. The initial angle for all systems was $\phi_0 = 90^\circ$.
	}
	\centering
\end{figure}

Concerning the characteristic time scales, at body temperature of $\SI{37}{\celsius} \approx \SI{310}{K}$ and for a radius of \SI{10}{nm} the Brownian relaxation time is about $\tau_\text{B} \approx \SI{2.6}{\mu s}$. This means that the Brownian relaxation limit is reached at $\approx \SI{400}{kHz}$. This value is in the range of the rotational relaxation limit.
The rotation limit $\tau_R$ varies stronger because it is field strength dependent but it is of a similar magnitude as the Brownian relaxation limit $\tau_B$ with a limit between $\SI{10}{\mu s}$ and $\SI{0.6}{\mu s}$.
The switching time $\tau_S$ in our model is about a thousand times shorter than the rotational relaxation time of the particle.
The N\'eel relaxation time for the magnetic core of radius \SI{9}{nm} for a temperature of \SI{310}{K} is $\tau_\text{n} \approx \SI{4}{s}$. This suggests high stability of the magnetic state and the prevention of the superparamagnetic behavior \cite{1959JAP....30S.120B}.
In a first comparison of the five different calculation models, the behavior of the orientation of magnetization is shown in Fig. \ref{fig:ana}.

\subsection{\label{sec:ana}Dynamics of a mechanical system}
Reducing our model to a system of mechanical equations leads to the rigid method. For the rigid$_\text{analytic}$ and rigid solution the results show periodic behavior but the particle's rotation is very stiff. The particle's movement is only rotating into the direction of the initial extension of the field and then back to its initial position, see Fig. \ref{fig:ana}. For high frequencies, much higher than the rotational relaxation limit, and low field strengths the hybrid method and analytical solution match well due to the limited rotation of the magnetization and easy axis and the reduced magnetic torque. The particle in the hybrid model for low field strengths will usually oscillate perpendicular to the field axis if the field strength is not too high.
This is only possible when internal magnetization dynamics are included.

At low frequencies, when the particle aligns with the field, it does so in an exponential fashion (see Eq. \eqref{eq:ana2}). A close alignment of the particle with the field leads to very small values for $\phi$ and can lead to mismatches of the analytic solution and in the time evolution of the rigid model due to numerical inaccuracies. 
Thus, the results match better at low fields when the particle is unable to completely relax via Brownian relaxation and align with the field. On the other hand, especially at higher frequencies and strong fields the N\'eel relaxation dominates for finite anisotropy and the results of the analytic solution would no longer suffice as an approximation.

\subsection{Dynamics of a solely magnetic system}\label{sec:mag}
This system only accounts for magnetic losses and is represented by the immobilized solution, where the easy axis is unable to rotate.
In Fig. \ref{fig:ana}, the magnetization is highly restricted in its motion because of the immobilized easy axis and the magnetization cannot deviate too much from the easy axis due to the high anisotropy constant.
In contrast to the Stoner-Wohlfarth-model, the angular velocity of the magnetic moment is finite and does not relax instantaneously. A mismatch of the results at high frequencies can be expected.

For an immobilized particle the magnetic losses are completely independent of the frequency until it starts to approach the magnetization switching limit $\tau_S$.
The energy output is the same as the particle will always switch at the same field strength leading to the same hysteretic losses. This system is not adjustable as it either dissipates no energy or it always outputs the same energy.

\subsection{Dynamics of the fully-coupled hybrid system}
\begin{figure}
	\includegraphics[width=0.8\columnwidth]{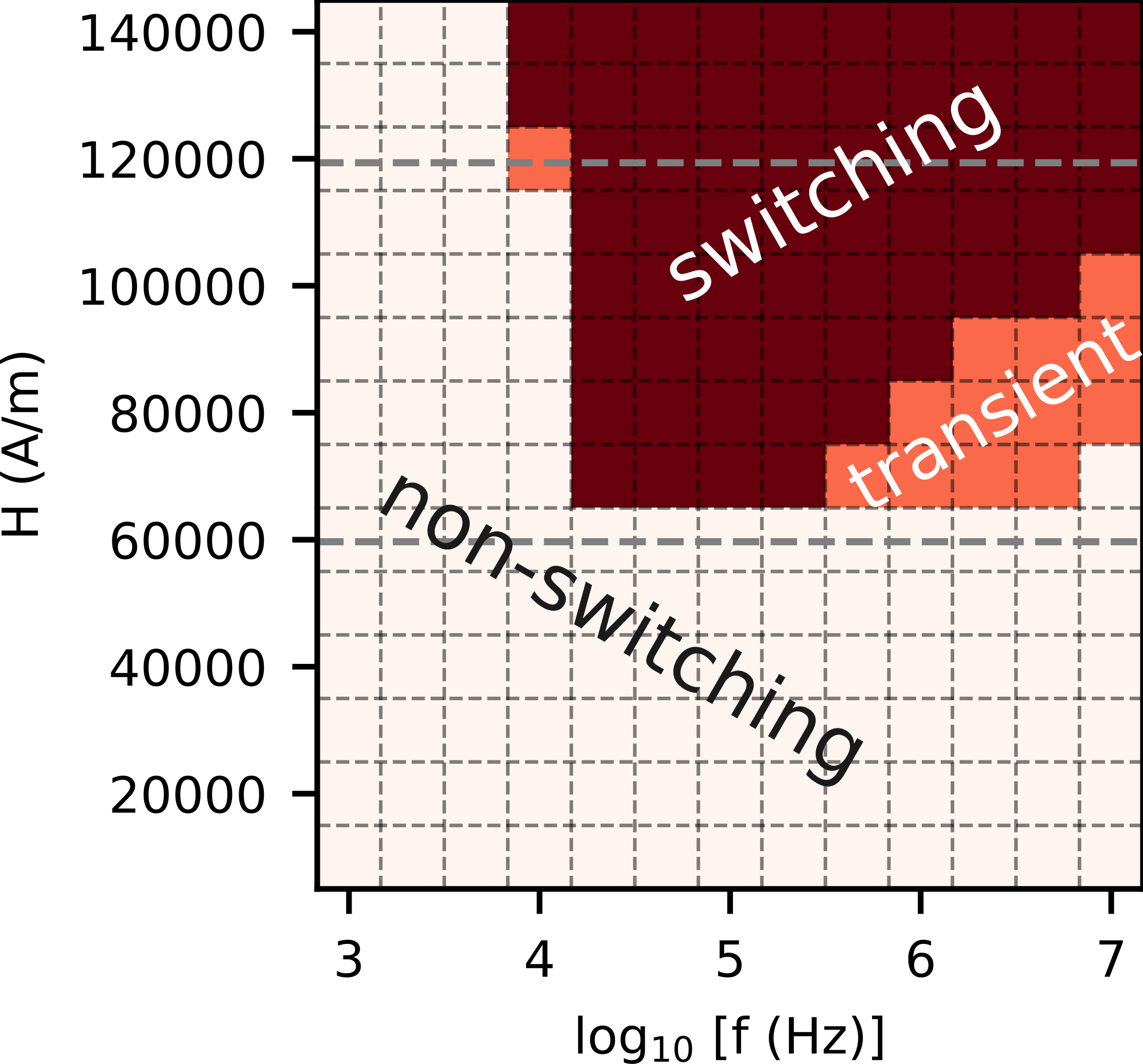}
	\caption{\label{fig:regions}Showing the region of magnetic reversal in the parameter space of field strength and frequency. In the dark region (\textit{switching}) the particle always settles into a switching steady state. This means the magnetization will continuously switch in the steady state according to the oscillation of the AMF. In the \textit{transient} region the steady state depends on the initial angle between easy axis and field. The particle rotates without the switching of the magnetization in the light region (\textit{non-switching}). The horizontal dashed lines indicate half ($\approx \SI{60}{kA/m}$) and the full critical field ($\approx \SI{120}{kA/m}$).
	}
\end{figure}

The two prior models show the behavior in the limit of no friction and high friction to the point of immobilization of the particle. Friction depends on the viscosity of the medium which determines the rotational relaxation time. Thus, the two models also represent the limits for frequencies well below and far above the Brownian relaxation time. Depending on the choice of material parameters and field strength, this leads to two distinct regions and a transient region of steady states, see Fig. \ref{fig:regions}.

The variety of steady states can be fundamentally subdivided into two categories: switching and non-switching. The regions indicate for which field parameters the magnetic moment flips or the particle rotates to accommodate for the change in field intensity.
The parameter space can be further divided by the critical field strength, half the critical field strength and the rotational relaxation limit for frequencies.

For an immobilized particle, the dissipated energy depends highly on the initial angle as predicted by the Stoner-Wohlfarth model but is generally independent of the frequency. N\'eel relaxation can also be influenced by the rotation of the particle. A rigid particle, on the other hand, gets limited strongly by the Brownian relaxation limit and thus has a lower energy output for higher frequencies.
Characteristics\scomment{Reviewer2 correction 27.7.2022} of both models will still appear in the hybrid method. Such as its two possible steady states that are quite similar to both previously mentioned models and their steady states.
The magnetization-switching behavior of the hybrid method is similar to the solely magnetic system, while the non-switching steady case is reminiscent of the rigid model.

In order to observe the particle transitioning to a steady state, where it reliably switches the orientation of magnetization, the Stoner-Wohlfarth model \cite{1948RSPTA.240..599S} can be utilized.
The Stoner-Wohlfarth astroid indicates that, depending on the alignment of field and easy axis, between half and the full critical field $H_{\text{crit}}$ (see Eq. \eqref{eq:crit}) strength has to be at least applied in order to switch the magnetization. The minimum switching field can switch the magnetization at an angle of exactly $45^\circ$ between the field and easy axis. Since the particle is also simultaneously rotating, this strict angle cannot be maintained. Thus, the particle starts switching its magnetization at a slightly higher field strength than half the critical field strength.
At around \SI{100}{kHz} the conditions are optimal for the magnetization to switch at the weakest field strength which is about \SI{65}{kA/m} or about $0.54 \; H_{\text{crit}}$.
Upon reaching the rotational relaxation limit the rotation of the easy axis is slowed down by the increasing viscous torque.
A similar critical field strength for the transition has also been found by Usadel \textit{et al.} \cite{doi:10.1063/1.4937919}.
The transition to the switching phenomena can thus be generally observed between \SI{65}{kA/m} and \SI{120}{kA/m}. In order to cover this transition region and also study the low field strength regime, a range from \SI{1}{kA/m} to \SI{140}{kA/m} is chosen.

As a side note, if easy axis and magnetization are perfectly aligned with the field, the magnetization will actually not start switching, because at that point the maximum for the critical field is necessary for switching and the system is perfectly balanced.
Due to thermal fluctuations this effect should not occur in experiments.
In the simulations the symmetry needs to be broken by very small perturbations to the initial angle.

\begin{figure}
	\includegraphics[width=0.8\columnwidth]{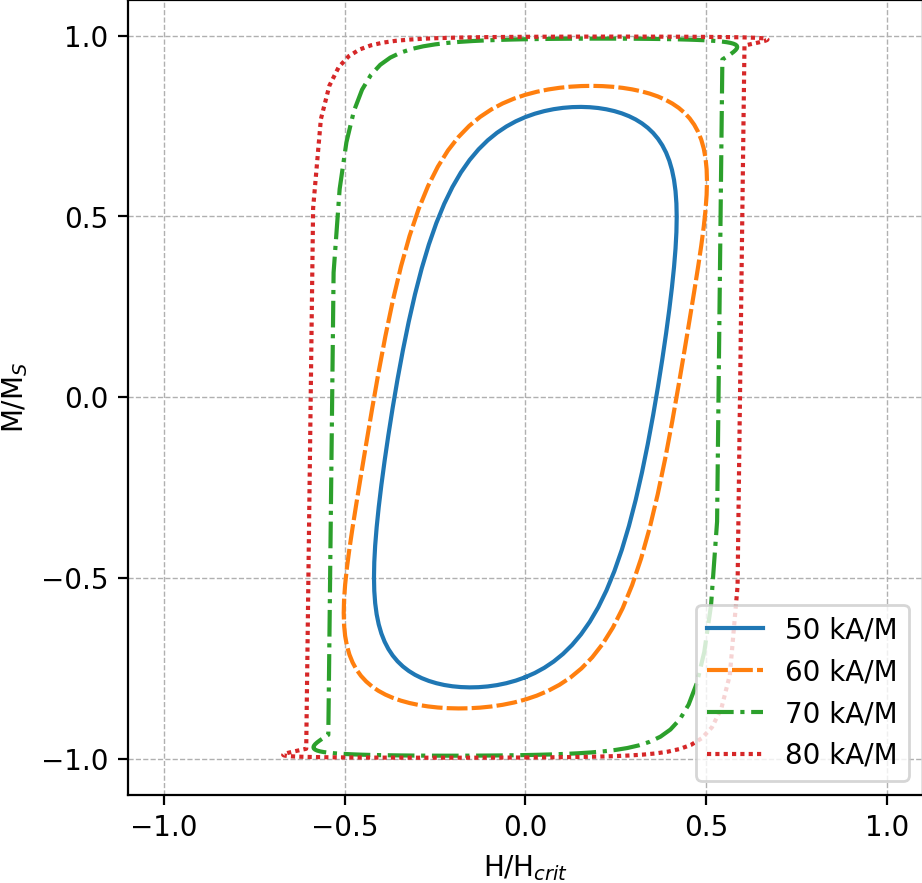}
	\caption{The hysteresis curve in the global frame of reference of an MNP during a steady state cycle of the AMF for four different field strengths (\SIrange{50}{80}{kA/m} from the inside to the outside) at \SI{500}{kHz}. The particle is simulated with the default parameters. The simulation includes mechanical rotation and therefore the curve captures both mechanical and magnetic processes. $M/M_s$ corresponds to the alignment of the magnetization with the field. Full saturation is reached when the magnetization is completely aligned with the field.}
	\label{fig:hyst}
\end{figure}

As previously mentioned, the hysteresis curve in the fixed coordinate system represents the total energetic losses due to N\'eel and Brownian relaxation during a cycle of the AMF. For field strengths below half the critical field strength $H_{\text{crit}}$ it is impossible for the magnetization to switch and thus only allows the particle to relax via Brownian relaxation. For intermediate field strengths, between half and the full critical field, it depends on the field frequency.
The hysteresis area seems to increase monotonically with the field strength but its shape seems to change as well.
If the magnetization does not switch, the shape is more rounded while at higher field strengths, due to the switching of the magnetization, the corners become sharp, see Fig. \ref{fig:hyst}. The switching leads to a sudden change in magnetization. Afterwards the particle continues to relax slightly via Brownian relaxation as well.
Although time is not resolved in the hysteresis loop, the difference in time scale for rotation (Eq. \eqref{eq:rot}) and magnetization switching (Eq. \eqref{eq:swi}) means that the particle spends significantly more time relaxing with Brownian relaxation, even though the switching of the magnetization is mostly responsible for the energy dissipation.
The area of the hysteresis loops in Fig. \ref{fig:hyst} increases abruptly when transitioning from Brownian to mostly N\'eel relaxation. For the lower field strengths the particle does not reach saturation $M/M_s$ which means that it is not able to fully relax via Brownian relaxation and the field is too weak to cause the magnetization to switch via N\'eel relaxation. For stronger fields the N\'eel relaxation might also not lead to a relaxed state at very high frequencies that approach the magnetization switching limit $1 /\tau_S$.

While magnetic losses can only occur if the magnetization switches, friction, although very limited, is always present. And even if no full relaxation is possible, rotation of the easy axis, albeit marginal, still occurs. The particle will settle in a steady state of the least energy losses and leads to an oscillation with small amplitude perpendicular to the field axis. 
For weak field strengths the easy axis remains almost motionless in its initial position. The magnetic torque in this region is not strong enough to significantly influence the easy axis during a cycle. This will be further discussed in section \ref{sec:ampl}.

The frequency of the AMF thus is also crucial in determining the behavior of the particle and relates to the time scales of the Brownian and N\'eel relaxation.
In the simulations a range between \SI{1}{kHz} to \SI{10}{MHz} is chosen because for medical applications the frequencies are rather low \cite{AtkinsonWilliamJ1984UFiH} but this range also shows the transition from Brownian to a dominating N\'eel relaxation and at higher frequencies also the emergence of a configuration that depends on the initial arrangement of easy axis and field (see "transient region" in Fig. \ref{fig:regions}).
Low frequencies allow for effective Brownian relaxation, where the easy axis can align with the field axis in positive and negative x-direction as the field alternates. On rare occasion the particles magnetization can switch for intermediate field strengths (\SIrange{60}{120}{kA/m}).
These spontaneous switching events may stem from the length of the time-step and could be avoided with a higher time resolution.
Increasing the frequency of the field closer to the Brownian relaxation limit, contrary to the previous case, sometimes the particle is too well aligned with the field such that the magnetization does not switch at all and remains motionless for one or more cycles of the field.
For slightly higher frequencies the previous inconsistencies vanish completely and for stronger field strengths the N\'eel relaxation becomes the preferred relaxation mechanism.

As the Brownian relaxation limit is approached, the frictional losses reach their maximum.
Although energy dissipation in the switching region is mainly dominated by N\'eel relaxation, at these frequencies there is also still a noticeable effect of Brownian relaxation. Energy losses due to magnetization switching are independent of the frequency, thus the additional friction contribution leads to the maximum in total energy losses per cycle.
Above the Brownian relaxation limit the losses due to friction drop off and for frequencies, for which the inverse approaches the magnetization switching time, the magnetization will be unable to completely relax but the frequencies in our simulations are lower than this threshold.
\begin{figure}
	\includegraphics[height=0.5\textheight]{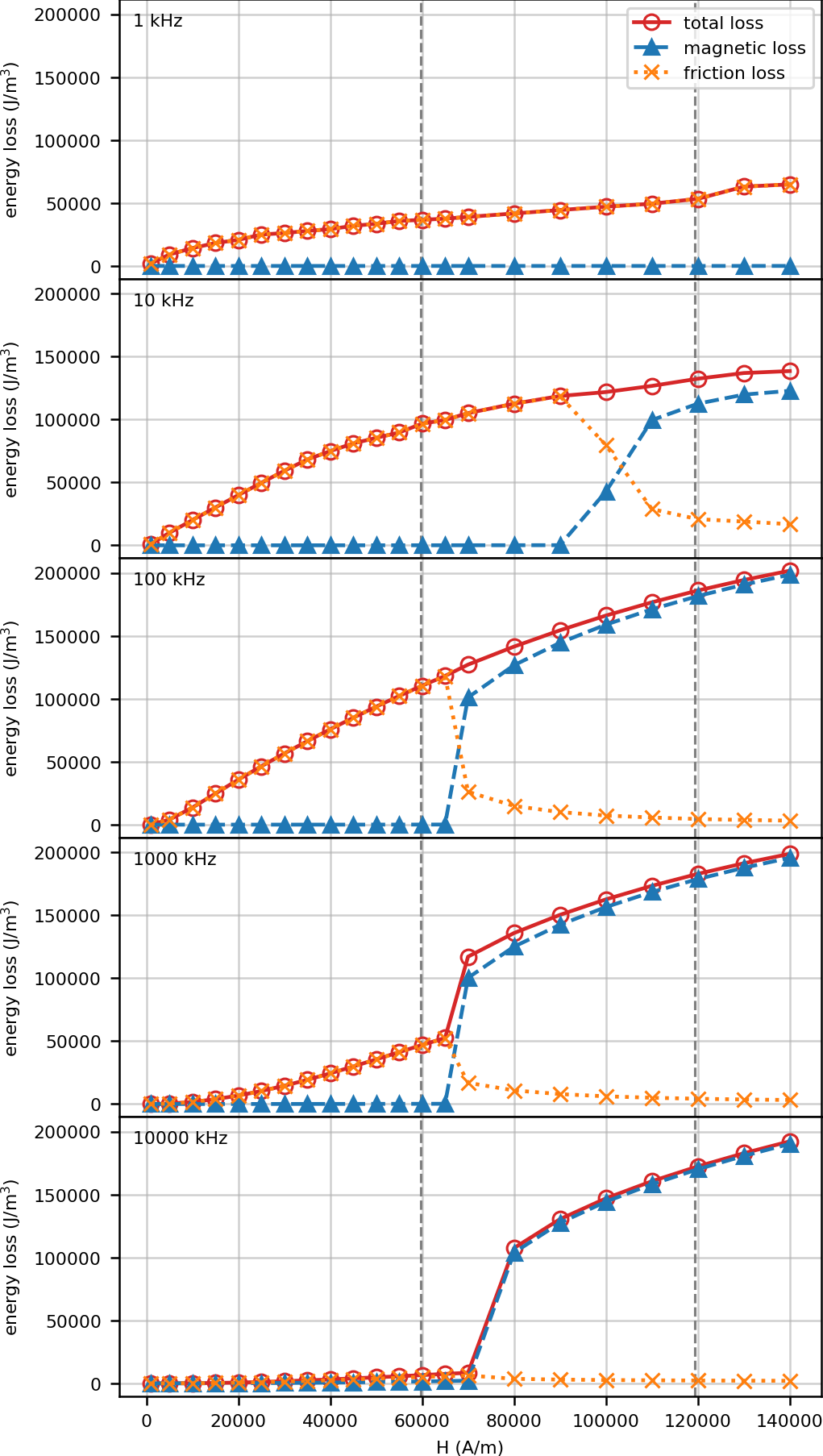}
	\caption{\label{fig:graph}The energy loss contributions of magnetic and friction processes for some sample of frequencies (increasing frequencies from top to bottom). At lower frequencies the accuracy of the different calculation methods for friction and total losses lead to some discrepancies where the friction losses occasionally appear to surpass the total losses.}
\end{figure}
\begin{figure*}
	\includegraphics[width=\textwidth]{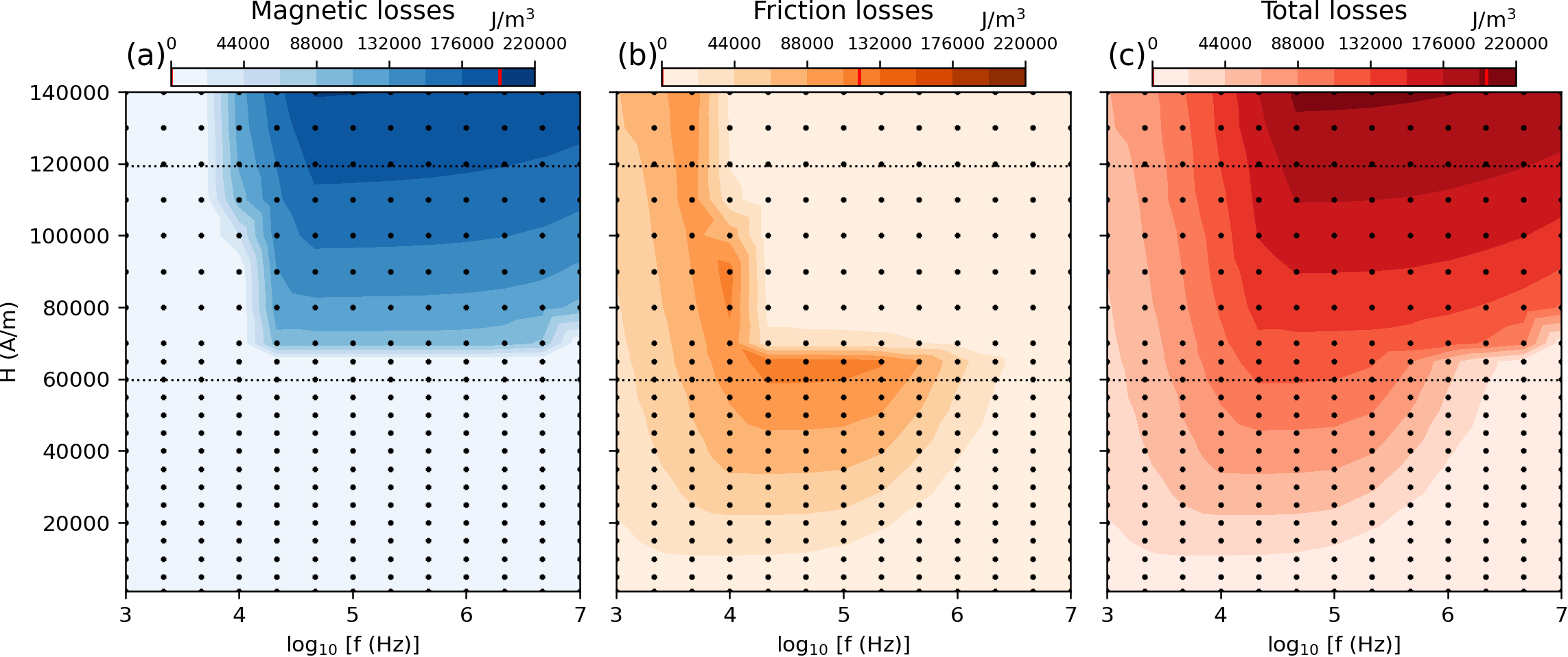}
	\caption{\label{fig:eloss}The energy losses as generated by magnetic processes (a), friction (b) and the total losses (c) in the space of the field parameters. The data points are logarithmically scaled according to the frequency. The values are taken for an initial angle $\phi_0 = 45^\circ$. The indicator lines in the colorbars show the maximum energy value. The color gradients follow the same scaling. Thin gray lines at \SI{60}{kA/m} and \SI{120}{kA/m} indicate half and full critical field strength respectively.}
\end{figure*}

These influences of frequency and field are depicted in Fig. \ref{fig:eloss}. In three plots of the space of field parameters, the magnetic (a), friction (b) and combined losses (c) are shown.

\begin{figure}
	\includegraphics[width=0.9\columnwidth]{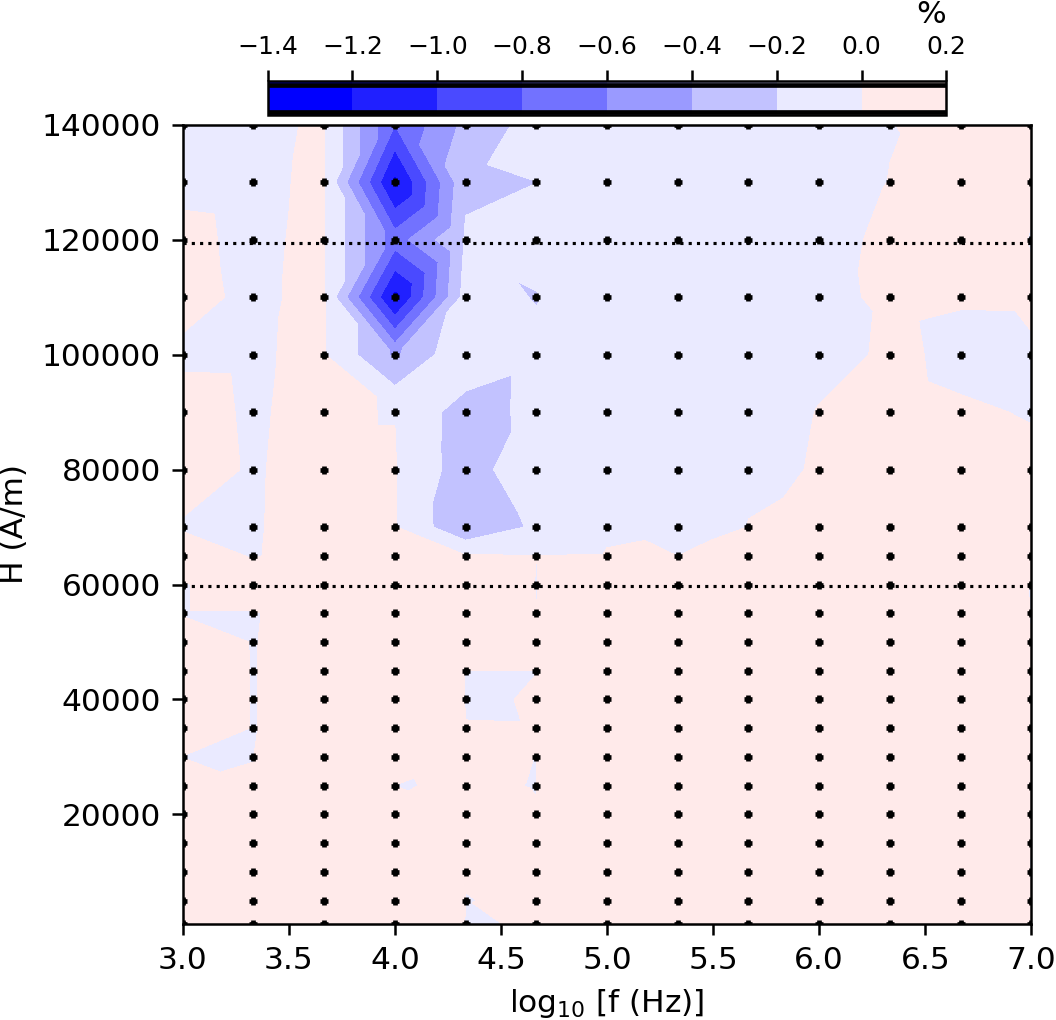}
	\caption{\label{fig:err}The difference of calculating the total energy loss per cycle directly via the hysteresis or as the sum of frictional and magnetic losses shown as a color gradient.
	These results are taken from simulations with an initial angle $\phi_0 = 45^\circ$. }
\end{figure}

Since the total losses can be computed in two ways, as explained in section \ref{sec:ene}, the difference is shown in Fig. \ref{fig:err}, where the difference is calculated by
\begin{equation}\label{eq:err}
    \Delta E = \frac{100}{E_\text{total}} ( E_\text{total} - (E_\text{mag} + E_\text{fric} )) .
\end{equation}

The difference in the calculation methods is generally very small ($< 1\%$).
The simulation results are evaluated for 10000 equidistant time steps per cycle in order to obtain sufficiently high accuracy.

The behavior of the particle is also influenced by material parameters. The important material parameters are the saturation magnetization, the anisotropy and the shape and size of the particle.
The saturation magnetization, together with the anisotropy constant, determines the critical field strength $H_\text{crit}$, see Eq. \eqref{eq:crit}.
The saturation magnetization and the field strength equally contribute to the calculation of $H_\text{crit}$ and thus the results for varying the field strength can be analogously applied to variation of the saturation magnetization.

The anisotropy energy is also a determining factor for the stability of the magnetization when exposed to thermal influences.
Fortunately, the anisotropy energy of the particle in the simulations is \SI{30}{kJ/m^3} and thus rather high. Together with the relatively large size of the particle, this results in high stability against thermal fluctuations and the superparamagnetic behavior is avoided.

Apart from the field and material parameters, the fluid also influences the dynamic behavior of the particle.
The fluid is primarily defined by its viscosity which limits the angular rotation of the particle and is a determining factor for friction and the Brownian relaxation time. Increasing the viscosity would lead to switching of the magnetic moment at lower frequencies or put in other words, it would shift the switching domain to lower frequencies, while increasing viscosity would approach the solely magnetic model discussed in section \ref{sec:mag}.

\begin{figure}
	\includegraphics[width=0.9\columnwidth]{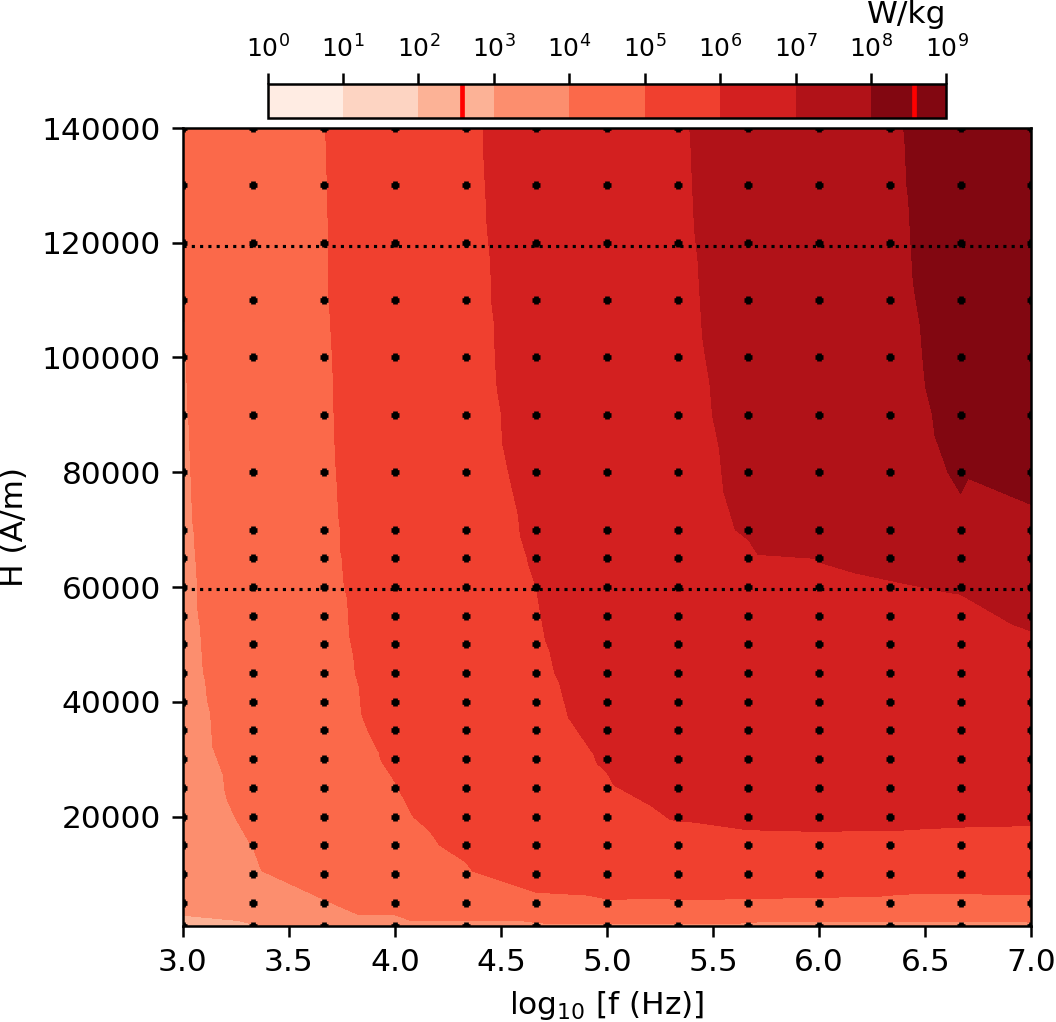}
	\caption{\label{fig:sar}
	The specific absorption rate (SAR) in a phase space diagram with logarithmic color scaling. The red lines in the colorbars indicate the minimum and maximum SAR value in the plot.
	The values are taken for an initial angle $\phi_0 = 45^\circ$.}
\end{figure}
The specific absorption rate SAR (see Fig. \ref{fig:sar}) is the most important result for hyperthermia in order to quantify the heating properties of the magnetic fluid and defined as the dissipated power per unit mass, see Eq. \eqref{eq:SAR}.
In general, the SAR increases with field strength and frequency. Noticeably for low field strengths ($< \SI{30}{kA/m}$) the SAR value drops off because of an insufficient magnetic torque and for frequencies above the Brownian relaxation limit the SAR value remains almost constant. 
For intermediate field strengths at frequencies above the Brownian relaxation limit, the reduced ability to rotate the easy axis leads to a jump in the SAR value at $\sim$ \SI{10}{MHz} and \SI{70}{kA/m} from low friction to high magnetic losses.

\subsection{Initial angle dependency - The transient region}\label{sec:trans}
Because of the dependency of the critical field strength $H_{\text{crit}}$ on the angle $\phi$ between easy axis and field in the Stoner-Wohlfarth model (see Fig. \ref{fig:astr}) we were interested to see if there are any differences in behavior of the particle depending on the initial angle. In general, every initial angle $\phi_0$ will result in the same steady state after a transient phase of a few cycles of the AMF. Since the motion of the particle is confined in a plane the results are also symmetric around the field axis.
During the cycles of the transient phase the particle can also change its behavior from switching to non-switching or vice-versa until it settles in the steady state.

If the field is slightly stronger than half the critical field strength $H_\text{crit}$ then, at frequencies exceeding the Brownian relaxation limit, two steady states can emerge for the same combination of field strength and frequency.
Depending on the initial angle the particle can either switch its magnetization or enter a niche steady state of oscillation shown in Fig. \ref{fig:regions} as "transient" region where the magnetization does not switch and the particle oscillates perpendicular to the field axis.
In case of obtuse angles ($90^\circ - 180^\circ$) the behavior is similar to simulations with lower field strengths without magnetization switching.
This is unusual since at intermediate field strength the steady state would be a state of switching of the magnetization for most initial angles. This range of angles inhibiting the magnetization switching widens at higher frequencies and for frequencies higher than $f > \SI{1}{MHz}$ completely blocks the particle from switching its magnetization at \SI{70}{kA/m}, which otherwise was possible at lower frequencies.

Since these simulations do not consider interactions between particles or the temperature, and therefore thermal activation of the magnetization switching, this second steady state might be too fragile to occur in experiments.
Previously balanced states could become less stable if temperature is included and the special case of steady states that depend on the initial angle could vanish.

\begin{figure}
	\includegraphics[width=0.9\columnwidth]{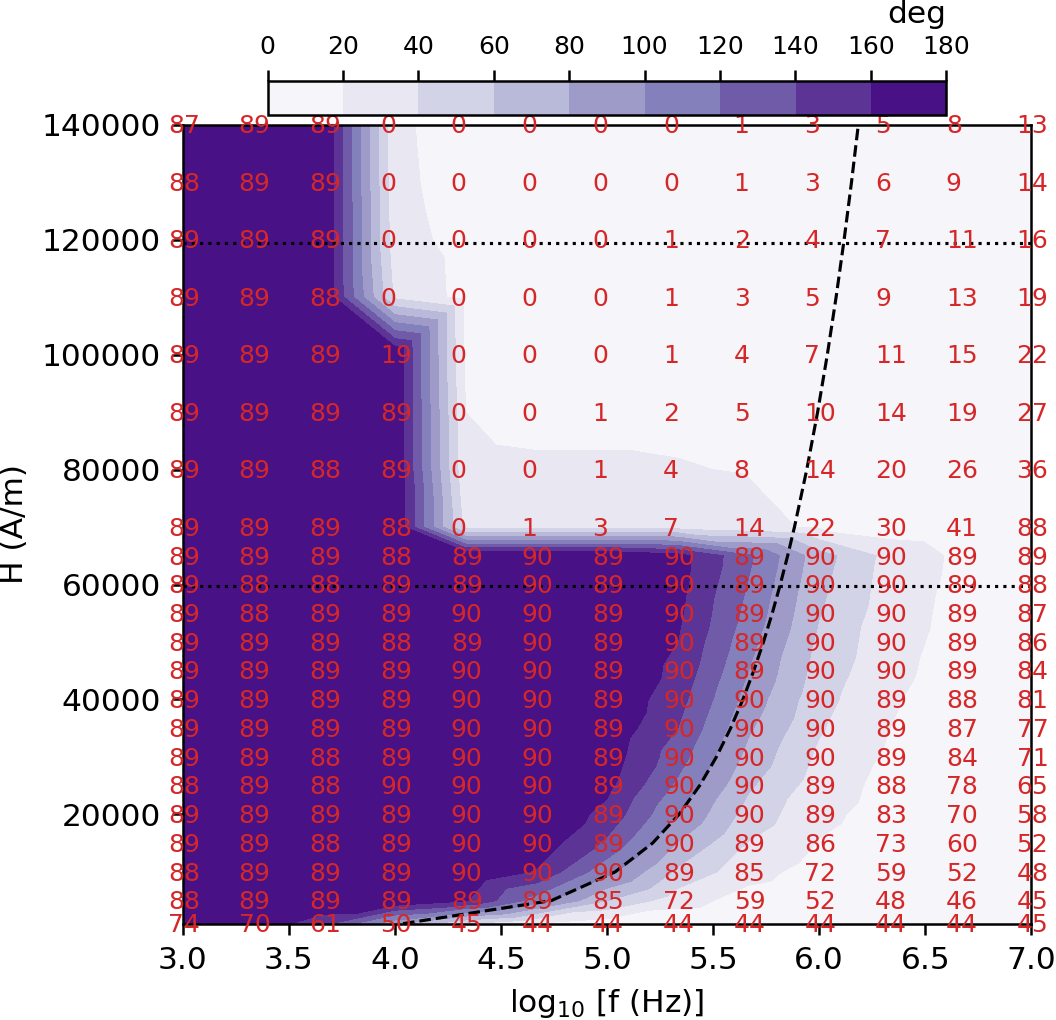}
	\caption{\label{fig:geom}The amplitude of the rotation of the easy axis as a gradient in degrees and the average angle of that oscillation written in numbers during one cycle of the AMF both given in degrees.
	The dashed curve indicates the rotational relaxation limit.
	The initial angle in the simulations is $\phi_0 = 45^\circ$.}
\end{figure}

At \SI{10}{kHz} and high field strengths the steady state is not completely stable and relaxation can alternate in the simulations between switching and non-switching of the magnetization contrary to the "transient"-region. Thus, at the boundary of the two regimes the steady state is not clearly defined and stands out in Fig. \ref{fig:regions} and \ref{fig:err}.

\subsection{\label{sec:ampl}Amplitude and average angle of oscillation}
At the nano-scale the inertial effects are negligible such that the easy axis cannot overshoot or perform a full rotation during one cycle of an AMF.
The maximum amplitude of the rotation of the easy axis is 180$^\circ$.
For low field strengths and frequencies above the rotational relaxation limit the amplitude shrinks drastically (see Fig. \ref{fig:geom}).magnetic
Although the particle can rotate for small field strengths, it turns very slowly and thus does not dissipate that much energy due to friction.

In general, if the particle's magnetization does not switch then it oscillates from 0$^\circ$ to 180$^\circ$ resulting in an average angle of 90$^\circ$, shown in numbers in Fig. \ref{fig:geom}) for low frequencies. The average angle remains the same even as the amplitude of oscillation decreases due to the rotational relaxation limit as the frequency increases. For very weak fields at high frequencies the magnetic torque is almost insufficient to even move the particle and the easy axis will remain close to its initial orientation, only oscillating slightly.

On the other hand, for fields exceeding the critical field strength the magnetization usually switches and the easy axis remains close to the field with a very small amplitude and an average angle close to 0$^\circ$.
This holds true in general for high field strengths, except for frequencies well below the Brownian relaxation limit, where the particle usually does not switch.

\section{Conclusion}\label{conclusion}
The framework developed in this work provides the foundation of an elaborate simulation model for magnetic fluids. By understanding the inner mechanisms of the particles in the fluid, more sophisticated predictions about the behavior of the fluid can be derived. Moreover, our model can show various interesting effects of a magnetic nanoparticle in a fluid, that simpler models cannot capture.
Although for specific cases simpler models may suffice, better results are usually obtained by the hybrid method, even though the computational cost is slightly higher.
With the mechanical motion derived from the torque equations acting on the particle and the LLG solving the magnetization dynamics inside of the particle, it has been shown for which configuration of field parameters the simulated particle settles in a switching or non-switching steady state. Another transient region in the space of field parameters in which the steady state depends on the initial angle between field and easy axis has also been observed and discussed.

Using the hysteresis curves to visualize the change and the underlying calculations for the energy losses clearly indicate the transition of the heating mechanisms and the total energy losses. The total dissipated energy increases continuously even as the system transitions from the non-switching to the switching steady state.
At higher frequencies this transition is more abrupt due to the rotational relaxation limit and the reduced frictional losses.
The magnetic hysteresis losses remain mostly independent of the frequency since they are caused by the irreversible switching of the magnetization, which depends mostly on the field strength. For that reason, the maximum energy losses are achieved when the Brownian relaxation can be maximized as Brownian relaxation is always present. This maximum total energy loss is reached shortly before reaching the rotational relaxation limit at around \SI{100}{kHz}.
The behavior of the easy axis and the magnetization have been thoroughly discussed and the results solidify this model as a model to study the physics of a magnetic fluid.

The SAR value has been found to increase monotonously with increasing field and frequency, however, for very low field strengths and frequencies over the rotational relaxation limit the SAR value remains almost constant.

Our future studies will focus on the further variation of the material and fluid parameters. This includes the shape of the particle which could introduce an additional anisotropy factor and would also change the viscous torque on the particle.
And although for the simulations the viscosity of water was chosen in this work, a more realistic fluid would be blood (at \SI{37}{\celsius}), which is more viscous.
Furthermore, thermal fluctuations not only influence the magnetization but can also change the size of the particle and the viscosity of the fluid and will be necessary to include in further studies.

\begin{acknowledgments}
	The authors wish to thank the "FWF - Der Wissenschaftsfonds" for funding under the project number P\,33748 and the Vienna Scientific Cluster (VSC) for providing the necessary computational resources.
	We acknowledge financial support by the Vienna Doctoral School in Physics (VDSP).
	PAS acknowledges support from the project "Computer modeling of magnetic nanosorbents", funded by the University of the Balearic Islands and the European Regional Development Fund.
	This research has been partially performed in the framework of the RSF Project No.19-12-00209.
\end{acknowledgments}

\appendix
\section{Characteristic time scales}
An important metric to estimate the influence of the mechanical and magnetic dynamics and their influence on the power loss is given by the characteristic relaxation time connected to the respective process.
The Brownian relaxation time $\tau_\text{B}$ is defined as the following
\begin{equation}\label{eq:brown}
	\tau_\text{B} = \frac{3 \eta V}{k_B T}.
\end{equation}
Here $\eta$ denotes the dynamic viscosity, $V$ is the hydrodynamic volume of the MNP, $k_B$ is the Boltzmann constant and $T$ is the temperature of the system.

Due to the zero-temperature approach in this work, an alternative description for the rotation limit can be derived from Eq. \eqref{eq:com} by rewriting $\dot{\phi} = \omega = 2 \pi / \tau_R$ and maximizing it. This yields the relaxation time
\begin{equation}\label{eq:rot}
	\tau_\text{R} = \frac{12 \pi V \eta }{\mu_0 M_S V_m H}
\end{equation}
which gives the limit for the rotation depending on the field strength.

The time it takes for the magnetization to switch its orientation via the LLG Eq. \eqref{eq:rllg} is given by \cite{kronmuller2007general}
\begin{equation}\label{eq:swi}
	\tau_{S} = \frac{2}{\gamma H} \frac{1 + \alpha^2}{\alpha}.
\end{equation}
The magnetization switching time is much shorter than the rotational relaxation limit.

The N\'eel relaxation time $\tau_\text{N}$ is the average lifetime of the magnetic state in absence of an external field and is derived from the N\'eel-Arrhenius law \cite{neel1949theorie}
\begin{equation}\label{eq:neel}
	\tau_\text{N} = \tau_0 \exp(\frac{K_u V_m}{k_B T})
\end{equation}
$\tau_0$ is the so-called attempt time and denotes the time frame that the magnetization should remain stable.
In this case, the inverse of the frequency of the AMF marks the attempt time since the particles magnetization should not switch due to thermal fluctuations for the duration of at least one cycle.
The relaxation time is defined by the ratio of the anisotropy energy of the particle $K_u V_m$ and the thermal energy $k_B T$.

\section{Magnetization switching field strength}
In the analysis of energy losses, the field strength necessary to switch the magnetization has to be considered. The astroid derived from the Stoner-Wohlfarth model \cite{1948RSPTA.240..599S} (see Fig. \ref{fig:astr}) allows to geometrically determine the orientation of the magnetic moment when an external magnetic field is applied. The astroid itself also indicates the field strength that is necessary to switch the magnetization in the particle. The required field strength to overcome the anisotropy barrier is also called critical field strength (compare to Eq. \eqref{eq:ani}):
\begin{equation}\label{eq:crit}
	H_\text{crit} = \frac{2 K_\text{u}}{\mu_0 M_\text{s}}
\end{equation}
Sharrock \cite{SharrockM.P1994Tdos} derived an elaborate method to analyse the thermal influence on the switching fields

\begin{equation}\label{eq:hc}
	H_S = H_\text{crit} \left \{ 1 - \left [ \left ( \frac{k_B T}{K_u V_m} \right ) \ln{\frac{t}{\tau_N} } \right ]^n \right \}
\end{equation}
where n is a factor dependent on the angle between the field and the easy axis. Setting the temperature $T$ to zero yields a coercive field same as the switching field of Stoner-Wohlfarth $H_\text{crit}$. Thus, in the limit of $\SI{0}{K}$ temperature our model holds true. For finite temperatures the here presented model overestimates the the critical switching field.

\bibliography{references}

\end{document}